\documentclass[preprint2]{aastex}
 
\usepackage[]{natbib}
\usepackage{color}

\slugcomment{}

\shorttitle{Non-thermal X-ray emission from HH 80}
\shortauthors{L\'opez-Santiago et al.}

\begin{document}


\title{Evidence of non-thermal X-ray emission from HH~80}


%
%
%
%
%
%
%

\author{J. L\'opez-Santiago\altaffilmark{1}, C. S. Peri\altaffilmark{2,3}, 
R. Bonito\altaffilmark{4,5}, M. Miceli\altaffilmark{5}, 
J. F. Albacete-Colombo\altaffilmark{6}, P. Benaglia\altaffilmark{2,3}, 
E. de Castro\altaffilmark{7}}


\altaffiltext{1}{Instituto de Matem\'atica Interdisciplinar, S. D. Astronom\'ia y Geodesia, 
Facultad de Ciencias Matem\'aticas, Universidad Complutense de Madrid, E-28040 Madrid, Spain}
\altaffiltext{2}{Instituto Argentino de Radioastronom\'ia (IAR), CCT La Plata (CONICET), C.C.5, 1894 
                      Villa Elisa, Buenos Aires, Argentina}
\altaffiltext{3}{Facultad de Ciencias Astron\'omicas y Geof\'isicas, Universidad Nacional de La Plata, 
                      Paseo del Bosque s/n, 1900 La Plata, Argentina}
\altaffiltext{4}{Dipartimento di Fisica e Chimica, Universit\`a di Palermo, Piazza del Parlamento 1, 90134 Palermo, Italy}
\altaffiltext{5}{INAF-Osservatorio Astronomico di Palermo, Piazza del Parlamento 1, I-90134 Palermo, Italy}
\altaffiltext{6}{Universidad Nacional del COMAHUE, 
                      Monse\~nor Esandi y Ayacucho, 8500 Viedma, R\'io Negro, Argentina}
\altaffiltext{7}{Dpto. de Astrof\'isica y CC. de la Atm\'osfera, Universidad Complutense de Madrid, E-28040 Madrid, Spain}


\begin{abstract}

Protostellar jets appear {at all stages of star formation when the accretion process 
is still at work. Jets travel at velocities} of hundreds of km~s$^{-1}$, creating strong shocks 
when interacting with interstellar medium. 
{Several cases of jets have been detected in X-rays, typically showing soft emission.} 
{For the first time, we report {evidence} of hard X-ray emission possibly related to non-thermal 
processes not explained by previous models of the post-shock emission predicted in the 
jet/ambient interaction scenario.}
HH~80 is {located at the south head} of the jet associated to the massive protostar 
IRAS~18162-2048. It shows soft and hard X-ray emission in regions that are spatially 
separated, with the soft X-ray emission region situated behind the region of hard X-ray 
emission. 
We propose a scenario for HH~80 where soft X-ray emission is associated to thermal 
processes from the interaction of the jet with {denser ambient matter} and the hard 
X-ray emission is produced by synchrotron radiation at the front shock. 
%

\end{abstract}





\section{Introduction}


Knots observed in the optical band (Herbig-Haro objects, HH) {originating from a YSO 
and traveling into the surrounding medium}  are  
associated to shock fronts and post-shock regions. The HH knotty structure along 
the jet axis, {typically} observed as a chain of knots, has been interpreted as pulses in 
the ejection of material by the star \citep[see review in][]{bon10a}, which can cause mutual 
interaction and collisions between knots \citep[an example of collision between optical 
knots is discussed in][for an X-ray emitting jet]{bon08}. 
{Hydrodynamic models predict X-ray emission by mechanical heating at the shock front, 
caused by the interaction between the jet and the ambient, with the shocked material 
reaching temperatures of 1~MK \citep{bon04}, thus emitting in the soft X-rays.}

Supersonic shock fronts and 
post-shock regions along the jet have been detected in a wide wavelength range, from 
radio to optical bands \citep[][and references therein]{rei01} and can be detected also in 
the ultraviolet {\citep{ort80,hol91,lis96,gom01,gom11,cof12,sch13}.} 
In recent years, several protostellar jets have been 
revealed also in X-rays \citep{pra01,pra04,fav02,fav06,bal03,kas05,gro06,gud08,bon10a,bon10b,bon11}. 


The Herbig-Haro complex HH 80-81 was detected in X-rays by \citet{pra04}. HH~80-81
is the south part of an extended jet whose {main source} is the high-mass protostar 
IRAS~18162-2048 \citep[$L = 17000\,L_\sun$;][]{car10}, {which is situated at a
distance of 1.7~kpc \citep{rod89} in the direction of the Sagittarius arm}. With a projected 
extension of $\simeq 5$~pc, this is one of the largest protostellar jets ever detected
\citep[see also][]{tsu04,rod05,car10}. 
\citet{pra04} distinguished two X-ray sources coincident with HH~80: a soft X-ray source 
($T \sim 1.5$~MK) related to HH~80A (the northern and {brighter} part of HH 80), and a hard source to the 
south of another H$_\alpha$ condensation (HH~80E) that they fitted with similar accuracy 
to both a high-temperature thermal model and a power-law with index $\sim 1$ {\citep[for 
a complete list of optical knots see][]{hea98}}. From the analysis 
of \emph{Chandra} spectra, the authors suggested that the hard X-ray source may be an 
embedded protostar, while the soft source is a region of shocked material 
{associated to the jet.}  

The HH~80-81 complex has been extensively studied at radio frequencies. \citet{mar93}
reported intensities and spectral {indices} for a number of radio bands using VLA 
observations. The authors indicated that radio emission from HH~80 and HH~81 is 
consistent with non-thermal radiation from relativistic electrons, 
according to measured spectral {indices}. More recently, \citet{car10} showed
that this jet is magnetized and linearly polarized. The latter is an indication of synchrotron 
radiation. The strength of the magnetic field {at the base of the jet} 
determined from the analysis of \citet{car10} is $B \approx 0.2$~mG. {Although  
there is no measurement of the field strength at the position of  HH~80 and HH~81, 
models of precessing jets for IRAS~18162-2048 reproduce well the position 
of these two knots and those detected to the North \citep{mas12}. This result has
been interpreted as a proof of the resemblance between the ejection mechanism in
both high-- and low--mass stars}.
 
Based on the most recent results from radio wavelengths {and our analysis of the X-ray 
properties of HH 80 detected with both XMM and Chandra (see Section~\ref{analysis})}, 
we propose a new scenario
for HH~80 in which the soft X-ray emission is produced in a shock front caused by 
the interaction of the jet with {the ambient medium}, while the hard X-ray emission is produced
by synchrotron emission of electrons in the jet magnetic field. In the following sections 
we show a multiwavelength analysis of HH~80 that supports our hypothesis.

\section{X-ray data analysis}
\label{analysis}

\subsection{\emph{XMM-Newton} observation} 
\label{xmm}

\begin{figure}[!t]
\centering
\includegraphics[width=\columnwidth,angle=0]{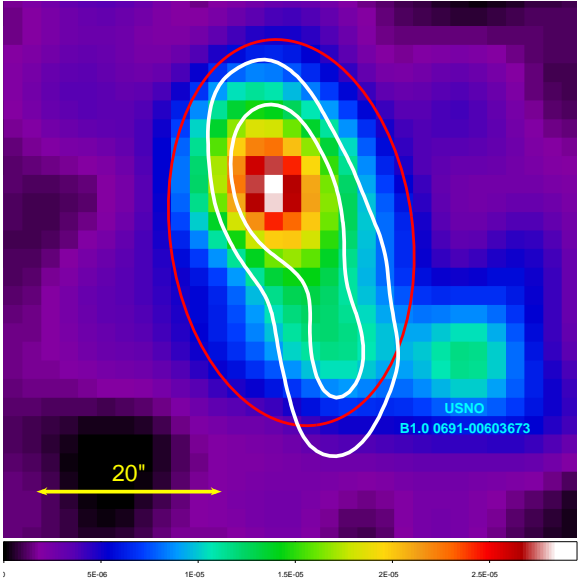}
\caption{
Count-rate image (MOS-equivalent counts per second per bin in the $0.3-1$ 
keV energy band). The bin size is $2''$, and the image is adaptively smoothed to a 
signal-to-noise ratio of ten. The region selected for the spectral analysis is shown in red. 
Radio contours from the data presented in \citet{mar93} are overplotted in white.
{North is up and East is to the left}.
\label{zoom}}
\end{figure}

We analyzed the \emph{XMM-Newton} EPIC archive observation ID 0149610401 
{(obs. date: 2003-09-14)} performed with the Medium filter by using the 
Full Frame Mode for the MOS {\citep{taa01}} and pn {\citep{sdb01}} 
cameras. {The observation was pointed towards the central source 
IRAS~18162-2048. With a telescope's field-of-view of 30$\arcmin$, this observation 
covers the whole jet, from HH~80N at the northeast to HH~80 and HH~81
at the southwest}.
We processed the data by using the Science Analysis System (SAS V12). 
Light curves, images, and spectra, were created by selecting events with 
PATTERN$\le$12 for the MOS cameras, PATTERN$\le$4 for the pn camera, and 
FLAG=0 for both. To reduce the contamination by soft proton flares, the event files 
were screened by adopting the sigma-clipping algorithm. 
The screened 
exposure times are $29$ ks, $30$ ks, and $23$ ks for MOS1, MOS2, and pn, respectively. 
Fig. \ref{zoom} shows the EPIC image of the source in the $0.5-1$ keV 
band. The image has been obtained  by superimposing (with the $EMOSAIC$ task) 
the MOS1, MOS2, and pn images, taking account of the differences between MOS and 
pn effective areas, and is background-subtracted, vignetting-corrected, and adaptively 
smoothed (with the task $ASMOOTH$).

\begin{figure}[!t]
\centering
\includegraphics[angle=-90,width=\columnwidth]{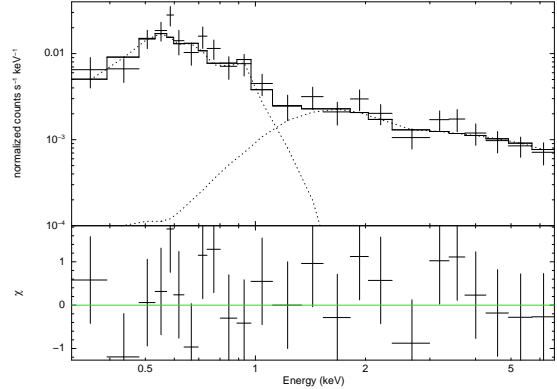}
\caption{PN spectrum of HH80  (extracted from the region shown in Fig. \ref{zoom}) 
with the corresponding APECT$+$Power-law best-fit model and residuals. The contribution 
of each component is shown.}
\label{fig:spec} 
\end{figure}

We performed the spectral analysis in the energy band $0.3-8$ keV using XSPEC 
V12.8. Spectra were extracted from the region shown {in Fig. \ref{zoom}} and 
rebinned to achieve a {signal-to-noise ratio per bin $>4$}. Spectral fittings were 
performed simultaneously on both MOS spectra and on the pn spectrum. The reported 
errors are at 90\% confidence. We verified that an isothermal model of optically thin 
plasma (APEC model in XPSEC) cannot describe the observed spectra ($\chi^2=260$ 
with 33 degrees of freedom, hereafter dof) and significantly underestimates the high-energy part of the 
spectrum showing systematic residuals above $\sim 1$ keV. We obtained a very good fit 
to the data ($\chi^2=29.1$ with 31 dof) by adding to the thermal model a power-law 
component. Figure \ref{fig:spec} shows the pn spectrum\footnote{Only the pn spectrum 
is shown for clarity.} with the best fit model and residuals. The thermal component, with 
temperature $kT_1=0.11\pm 0.03$ keV well {describes} the low energy part of the spectrum, 
while contribution of the power-law component, which shows a very flat spectral index 
$\Gamma=0.8^{+0.3}_{-0.2}$, dominates above $\sim 1.5$ keV. We also adopted a 
model with two APEC components to investigate a possible thermal origin for the hard 
X-ray emission. Though this model provides an acceptable fit to the spectra ($\chi^2=33.7$ 
with 31 dof), it provides an unrealistically high value for the temperature of the hot plasma, 
$kT_2>27$ keV. Therefore, we conclude that the hard X-ray emission does not have a 
thermal origin. 

\subsection{\emph{Chandra} observation}
\label{chandra}

HH~80 was observed twice with the {Advanced CCD Imaging Spectrometer (ACIS)} 
onboard the \emph{Chandra} X-ray observatory (Obs. Ids. 2535 and 6405).
HH~80  is off-axis in the latter and so, those data were not used for our study. 
During the observation performed in 2002, data were acquired in faint
mode for a total of 37.7~ks.

Data reduction, starting with the level 1 event list provided by the pipeline processing 
tool of \emph{Chandra}, was carried out with CIAO 4.4 and the CALDB 4.5.3 set of 
calibration files. We produced a level 2 event file using the \emph{chandra\_repro}
CIAO task, taking advantage of the F-mode enhanced background, filtering and 
retaining only events with $grades = 0, 2, 3, 4, 6$ and $status = 0$.  Since the images 
of the ACIS-S4 chip show a variable pattern of linear streaks, we created a new 
bad-pixel file to improve the streak detection efficiency in S4 chip. 
%
{We filtered events in the 0.5-8 keV range. By performing a simultaneous spectral 
fitting of the Chandra dataset with the XMM dataset, we derived compatible results 
as described in Section~\ref{xmm}.}

\begin{figure}[!t]
\centering
\includegraphics[width=0.45\columnwidth]{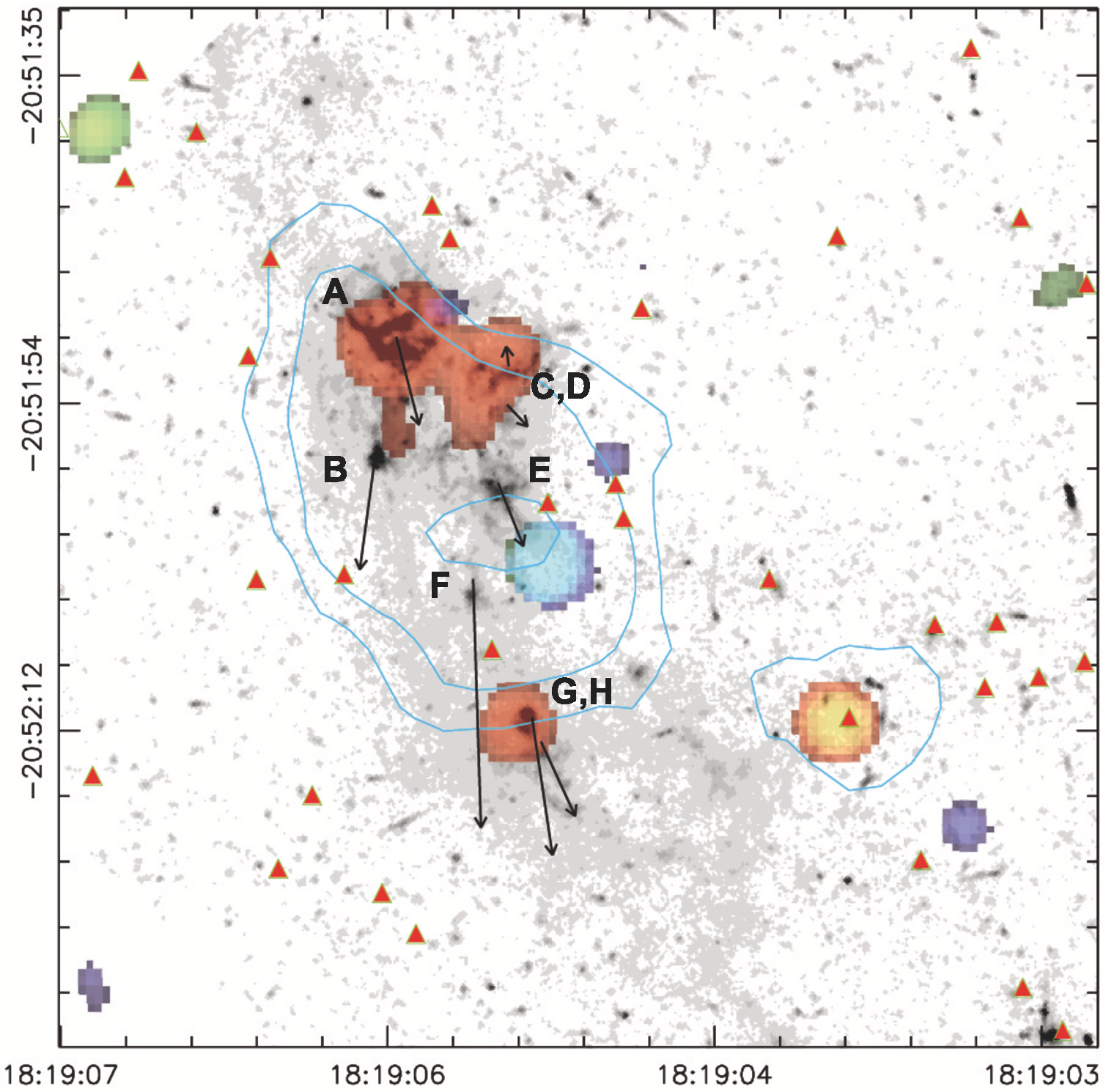}
\includegraphics[width=0.45\columnwidth]{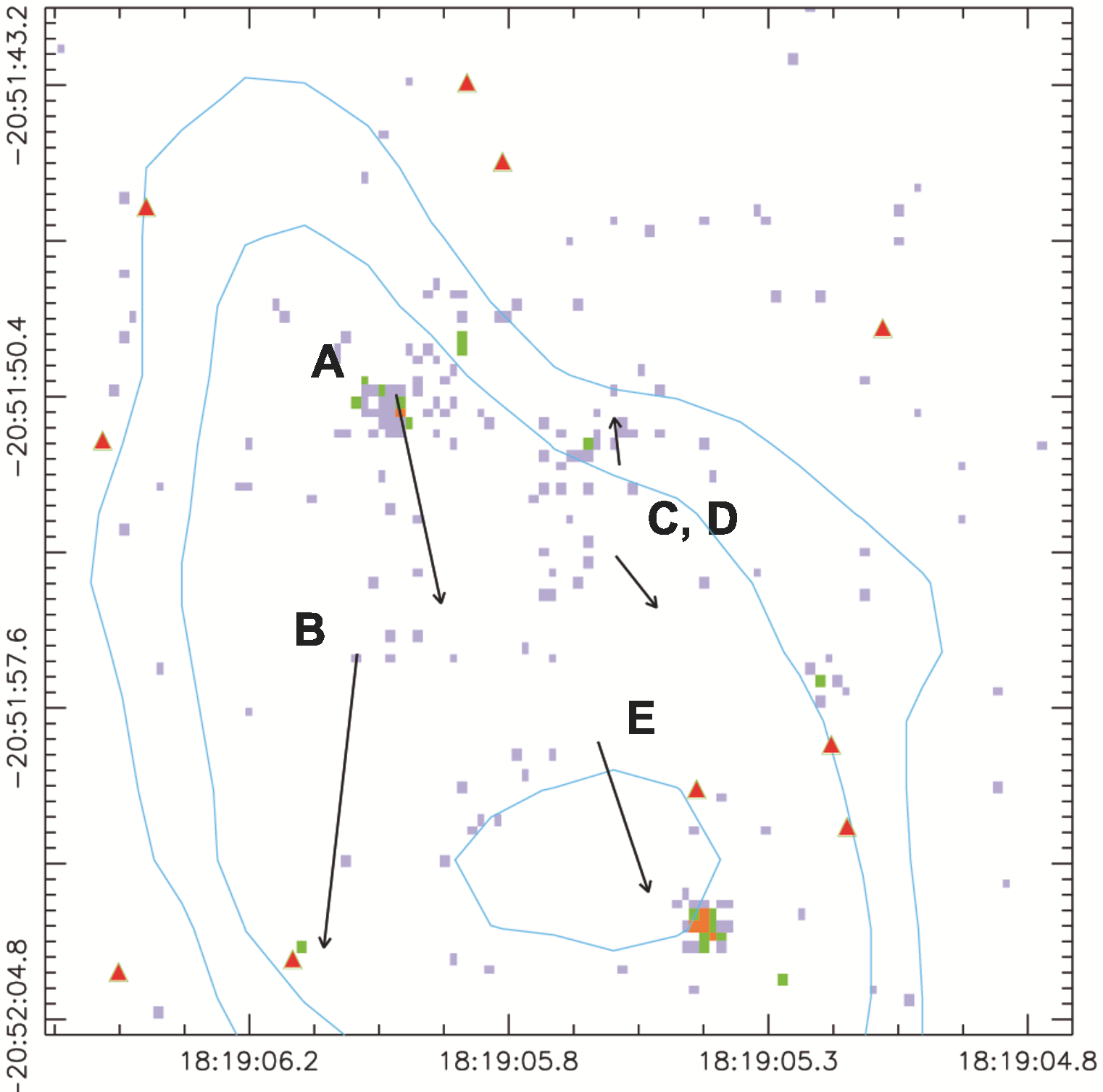}
\caption{
\textbf{Left:} WPC2 H$_\alpha$ image of HH~80 with \emph{XMM-Newton} EPIC-pn contours
and \emph{Chandra} smoothed image overplotted in false color: red for 0.3--1.2~keV, green for 
1.2--2.5~keV and blue for 2.5--4.5~keV.  {Coordinates are J2000.}
Proper motions are from \citet{hea98}, {transformed 
to J2000}. Filled triangles are \emph{Spitzer} sources in the [3.6]~$\mu$m band of IRAC with $S/N \ge 2$.
\textbf{Right: central region of HH~80. This is a refined \emph{Chandra} image 
resulting from the sub-pixel repositioning algorithm.}
The binning is 0.5. Contours, proper motions, triangles and flags are the same than in the left panel.  
\label{chandra}}
\end{figure}

{Chandra/ACIS resolution allowed us to distinguish the soft and hard X-ray emitting 
components (see Fig.~\ref{chandra}). In order to analyze the morphology of these 
two components of the HH 80 complex structure, we also applied the sub-pixel 
repositioning algorithm available in CIAO (EDSER) to the Chandra images to refine the event 
positions \citep{li04}. 
The improved image can therefore be resampled at one-half of the native ACIS pixel scale 
(0.5/2\arcsec). {Both the soft and the hard components show a hint of elongation which 
appears, if any, aligned} with the jet axis and the proper motion of the E knot (see Fig.~\ref{chandra}, 
right panel). Note that recently, an asymmetry in the Chandra 
point spread function (PSF) has been discovered located at position angle P.A. = 195 - roll angle 
$\pm 25$~deg, corresponding to P.A. = 285.844~deg which is not coincident with the direction of the 
extension of the X-ray sources nor the jet axis.}

\section{Search for optical and infrared counterparts}
\label{cross}

We looked for counterparts to the different X-ray sources associated to HH~80
at other wavelengths, with the aim of discarding an extragalactic origin for the X-ray 
emission in this region. We analyzed data from the \emph{Spitzer} mission 
and inspected the H$_\alpha$ image taken with the WFPC2 onboard the \emph{Hubble} 
Space Telescope \citep{hea98}. From Fig.~\ref{chandra}, it seems clear that there 
is no {point-like} source associated to the hard X-ray source {(below knot E)}, 
while the soft X-ray sources are related to different H$_\alpha$ knots of the jet
{(precisely knots A-D).}

The \emph{Spitzer} Heritage Archive at the NASA, IPAC Infrared Science Archive
contains one observation from the GLIMPSE project covering the region 
of HH~80-81 (AORKEY 21289984) and a pointed observation in map mode 
with both the IRAC, Infrared Array Camera (AORKEY 11069952) and the MIPS, 
Multiband Imaging Photometry for SIRTF (AORKEY 11072256). With a 
configuration with frame time 10.4 s, the pointed observation of HH~80-81 is 
substantially deeper, reaching limiting magnitudes in the [3.6]~$\mu$m band 
$\simeq 20$~mag for sources detected above $S/N = 2$ ($F \simeq 30$~$\mu$Jy). 
We followed the procedure
described in \citet{lop13} to find sources in this \emph{Spitzer} observation and 
performed aperture photometry. Both panels in Fig.~\ref{chandra} show the infrared sources 
detected by us in the proximities of HH~80 (filled triangles). The only infrared 
source close to the hard X-ray source is located $> 5 \arcsec$ to the North 
of the \emph{Chandra} source and it is detected only in the [3.6]~$\mu$m band
with $S/N \simeq 2$. 
{Given this low IR-signal and the size of the angular offset, any physical association is unlikely.}

\section{Discussion}

HH~80-81 is the head of a jet whose origin is the massive protostar 
IRAS~18162-2048. The jet emits at radio frequencies \citep{mar93} and this 
emission is linearly polarized \citep{car10}. The region of HH~80
shows radio emission with negative spectral 
index $\alpha = -0.3$ \citep{mar93}, indicating that it is dominated by non-thermal 
radiation \citep[e.g.][]{gin64}.
In addition to HH~80-81, only a few young stellar objects show {evidence}
of non-thermal radio emission in their outflows: Serpens triple radio source, 
Cepheus A East sources 1-7, W3(H$_2$O), L778 VLA~5 and IRAS~16547-4247 
\citep[see][and references therein]{car10}. Radio-synchrotron polarized emission 
has been detected only in IRAS~18162-2048.

The HH~80 \emph{XMM-Newton} spectrum shows a soft component, which is compatible 
with thermal plasma emission at a temperature $\sim 10^6$~K, plus a hard component
with a power-law spectrum with index $\alpha \simeq 0.8$. 
\emph{Chandra} is able to separate both spectral components spatially. The soft 
component is correlated with the denser H$_\alpha$ region (HH~80A, C and D). The hard 
component is located just in front of HH~80E in a region with low gas density
(see Fig.~\ref{chandra}). 

{\citet{bon10b} results suggest that HH80/81 and probably also the X-ray emitting jet discovered 
by \citet{tsu04} {(TKH8)}, for which high velocities are needed to explain their spectra, are not compatible 
with the other soft X-ray emitting jets well modeled by the post-shock emission due to the 
jet/ambient interaction \citep[as in the wide exploration of the parameter space performed 
also in][]{bon07}. In Fig. 7 of \citet{bon10b}, in fact, both HH 80 and TKH8 cannot be described 
by the model of the emission from the post-shock formed by the jet/ambient interaction as 
in the cases of the other jets emitting soft X-rays; furthermore, the shock velocity deduced in 
these two jets is high with respect to the other soft jets.}

{The analysis of the morphology of the {\emph{Chandra}/ACIS} detected source at sub-pixel 
resolution shows a hint of elongation of the hard component along the jet axis and the proper 
motion of the E knot (note that the PSF asymmetry has been computed and is not aligned 
with the jet axis but, on the contrary, it is almost perpendicular to the proper motion of the knots 
within the jet). Available Chandra/ACIS data allowed us to discriminate the two X-ray emitting 
components detected in HH 80: the soft component is almost coincident with the A knot, and 
the hard component is located along the jet axis further away with respect to the E knot and 
along its proper motion (see Fig.~\ref{chandra}, right panel). A deeper Chandra observation 
could help in distinguishing the hard component morphology and in 
performing a spatially resolved spectral analysis of the different sources. Also with a time 
baseline of more than ten years, it will be possible to measure the proper motion of the source, 
if any, and compare with radio proper motions.}

The hard X-ray source has been previously proposed to be an embedded 
protostar not related to the jet itself \citep[see][]{pra04}. 
To explore this scenario, we looked for counterparts to the hard X-ray source 
in any photometric band, from optical to infrared. As we mentioned in Section~\ref{cross}, 
no counterpart was found down to 20 mag in the \emph{Spitzer}/IRAC 
bands, neither in MIPS data. This result likely discards the scenario of the
embedded protostar. 

The possible relation of the  hard X-ray source with a highly obscured AGN {appears 
also unlikely}. With a measured (unabsorbed) X-ray flux in the [0.3-10]~keV energy 
band of $f_\mathrm{X} \sim 10^{-14}$~erg\,cm$^{-2}$\,s$^{-1}$, the source is quite 
bright in X-rays. According to the $\log N$-$\log S$ distribution of bright AGN 
\citep[e.g.][]{cec04}, the density of these objects at this flux is 
$\sim 100$ deg$^{-2}$. Following \citet{com11}, the probability that an AGN were 
detected inside the \emph{XMM-Newton} contours is $P \sim 0.15$ (15\%) and the 
probability that it were situated at less than $5 \arcsec$ of the hard X-ray source 
is $P < 0.02$ (2\%). 

At a distance of 1.7~kpc, the X-ray luminosity of the hard X-ray source is 
$L_\mathrm{X} \simeq 3 \times 10^{30}$~erg\,s$^{-1}$ 
{($f_\mathrm{X} \sim 10^{-14}$~erg\,cm$^{-2}$\,s$^{-1}$ for the hard X-ray source).
This is very close} to the value expected by \citet{bos10} for secondary electrons and positrons from 
proton-proton collisions \citep[see their Fig.~5 and][]{ara12}. {Note that 
the authors considered a particle density $n_c \sim 10^2 - 10^3$~cm$^{-3}$, 
bolometric luminosity $L_\star \simeq 2 \times 10^4$~L$_\odot$ and jet velocity at the base
$v_j \sim 10^3$~km~s$^{-1}$ \citep[see Table~1 of][for a complete list of observed and derived 
parameters for HH~80]{bos10}. The magnetic field determined by the authors at the position 
of HH~80 is $B \sim 0.02$~mG. This is one order of magnitude lower than the magnetic field
strength determined by \citet{car10} at the base of the jet but consistent with the value expected 
at the distance of HH~80 from the star \citep{bos10}}.

\section{Conclusions}

Supported by the results of our analysis of X-ray \emph{Chandra} and \emph{XMM-Newton} 
data and cross-correlation with optical and infrared databases, we propose a new
scenario for the nature of the hard X-ray emission detected at the edge of HH~80. 
While the soft X-ray emission is produced in the post-shock region after interaction of the 
jet with dense medium, the hard X-ray emission is produced by synchrotron radiation 
of accelerated particles at the front-shock. The spectral index of the power-law spectrum 
in both X-ray and radio wavelengths is consistent with this hypothesis. In addition, the X-ray 
luminosity assuming a distance of 1.7~kpc agrees well with the predictions by models of 
non-thermal emission for HH~80. This is the first evidence of non-thermal emission 
detected in X-rays for a stellar jet.

\acknowledgments

This work was supported by the Spanish Government  (AYA2011-29754-C03-01 
and AYA2011-29754-C03-03). C.~S.~P. was granted by the COSPAR Capacity Building 
Workshop Fellowship Program {and the Universidad Nacional de La Plata} 
to develop this work at the Universidad Complutense de Madrid. 
M.~M. is supported through the ASI-INAF contract I$/$009$/$10$/$0. 
{JFAC is a staff researcher of the CONICET and acknowledges support by
grant PIP~2011~0100285.}
J.~L.-S. would like to thank Drs. F.~J. Carrera, P.~G. P\'erez-Gonz\'alez 
and S. Mateos for fruitful discussion on spectral X-ray emission properties of AGN.
{We also acknowledges the referee of this manuscript for his/her suggestions 
and comments.}



{\it Facilities:} \facility{XMM-Newton (EPIC)}. \facility{Chandra (ACIS)}. 
\facility{Spitzer (IRAC, MIPS)}.

\clearpage

\end{document}